\documentclass[sigconf]{acmart}

\usepackage{booktabs} 

\copyrightyear{2020} 
\acmYear{2020} 
\setcopyright{acmcopyright}
\acmConference[WSDM '20]{The Thirteenth ACM International Conference on Web Search and Data Mining}{February 3--7, 2020}{Houston, TX, USA}
\acmBooktitle{The Thirteenth ACM International Conference on Web Search and Data Mining (WSDM '20), February 3--7, 2020, Houston, TX, USA}
\acmPrice{15.00}
\acmDOI{10.1145/3336191.3371842}
\acmISBN{978-1-4503-6822-3/20/02}

\usepackage{amssymb}
\usepackage{amsmath}
\usepackage{amscd,amsfonts,amsbsy,rotating}
\usepackage{balance}
\usepackage{graphicx}
\usepackage{epsfig,epstopdf}
\usepackage{subfigure}
\usepackage{multirow}
\usepackage{booktabs}
\usepackage{color,xcolor}
\usepackage{url}
\usepackage{latexsym,bm}
\usepackage{enumitem,balance,mathtools}
\usepackage{wrapfig}
\usepackage{euscript}
\usepackage{algorithm}
\usepackage{algorithmic}
\usepackage{ifpdf}
\usepackage{diagbox}
\usepackage{caption}
\usepackage{makecell}
\usepackage{subfigure}
\usepackage{bm}

\newcommand{\bs}{\boldsymbol}

\newcommand{\bh}{\bs{h}}
\newcommand{\btheta}{\bs{\theta}}

\newcommand{\score}{{SCoRe}}
\newcommand{\mg}{\bm{G}}

\DeclareMathOperator*{\argmin}{arg\,min}

\usepackage{array}
\newcolumntype{N}{@{}m{0pt}@{}}

\begin{document}

\title{Sequential Recommendation with Dual Side \\Neighbor-based Collaborative Relation Modeling}

\author{Jiarui Qin$^1$, Kan Ren$^2$, Yuchen Fang$^1$, Weinan Zhang$^1$, Yong Yu$^1$}
\affiliation{
  \institution{$^1$Shanghai Jiao Tong University, Shanghai, China\\
  $^2$Microsoft Research Asia, Beijing, China\\
    \{qinjr, arthur\_fyc, wnzhang, yyu\}@apex.sjtu.edu.cn, kan.ren@microsoft.com\\~}
}

\fancyhead{}
\renewcommand{\shortauthors}{J. Qin, et al.}
\renewcommand{\shorttitle}{Sequential Collaborative Recommender}
\settopmatter{printacmref=false}

\begin{abstract}
Sequential recommendation task aims to predict user preference over items in the future given user historical behaviors.
The order of user behaviors implies that there are resourceful sequential patterns embedded in the behavior history which reveal the underlying dynamics of user interests. 
Various sequential recommendation methods are proposed to model the dynamic user behaviors. However, most of the models only consider the user's own behaviors and dynamics, while ignoring the collaborative relations among users and items, i.e., similar tastes of users or analogous properties of items. Without modeling collaborative relations, those methods suffer from the lack of recommendation diversity and thus may have worse performance.
Worse still, most existing methods only consider the user-side sequence and ignore the temporal dynamics on the item side.
To tackle the problems of the current sequential recommendation models, we propose \textbf{S}equential \textbf{Co}llaborative \textbf{Re}commender (\score) which effectively mines high-order collaborative information using cross-neighbor relation modeling and, additionally utilizes both user-side and item-side historical sequences to better capture user and item dynamics. Experiments on three real-world yet large-scale datasets demonstrate the superiority of the proposed model over strong baselines.
\end{abstract}


\settopmatter{printacmref=false} 

\maketitle

{\fontsize{8pt}{8pt} \selectfont
	\textbf{ACM Reference Format:}\\
	Jiarui Qin, Kan Ren, Yuchen Fang, Weinan Zhang, Yong Yu. 2020. Sequential Recommendation with Dual Side Neighbor-based Collaborative Relation Modeling. In \textit{Proceedings of the 13th ACM International Conference on Web Search and Data Mining (WSDM '20), February 3-7, 2020, Houston, TX, USA.} ACM, New York, NY, USA, 9 pages. https://doi.org/10.1145/3336191.3371842 }
	
\section{Introduction}

With the emergence of large online information systems such as e-commerce platform, the amount of user behavioral data grows rapidly. Therefore, in recent years, the researchers in both academic and industrial fields have devoted many efforts on sequential recommendation task which aims to mine the resourceful yet complex temporal dynamics embedded in user behavior sequences.

As has been stated in many related works \cite{hidasi2017recurrent,koren2009collaborative,he2016vista,agarwal2009spatio}, the temporal dynamics have high impacts on the future user behaviors, especially accounted for concept drifting \cite{widmer1996learning}, long-term behavior dependency \cite{koren2009collaborative}, periodic patterns \cite{ren2018repeatnet}, etc.

Commonly, the current sequential recommendation models regard user's purchasing or browsing behaviors as ``token''s in the natural language processing (NLP) field. And the mainstream models use sequential modeling techniques that are widely used in NLP such as recurrent neural networks (RNNs) \cite{hidasi2015session,zhou2018deepb}, convolutional neural networks (CNNs) \cite{tang2018personalized} and Transformer \cite{kang2018self}. These models make huge success on sequential recommendation task with many deployed real-world applications \cite{zhou2018deepb,wu2016personal}.

Despite the success of current sequential recommendation methods, there are still some limitations of them. The first one is that most of the models \cite{hidasi2015session,tang2018personalized,kang2018self} only consider user's (or item's) own interaction history, while ignoring similar users or items that have collaborative relations with itself. 
Therefore, each user (item) only know her (its) own behaviors, it is bad for the variety of recommendation and may hurt the recommendation performance.

\begin{figure}[h]
	\centering
	\includegraphics[width=1.0\columnwidth]{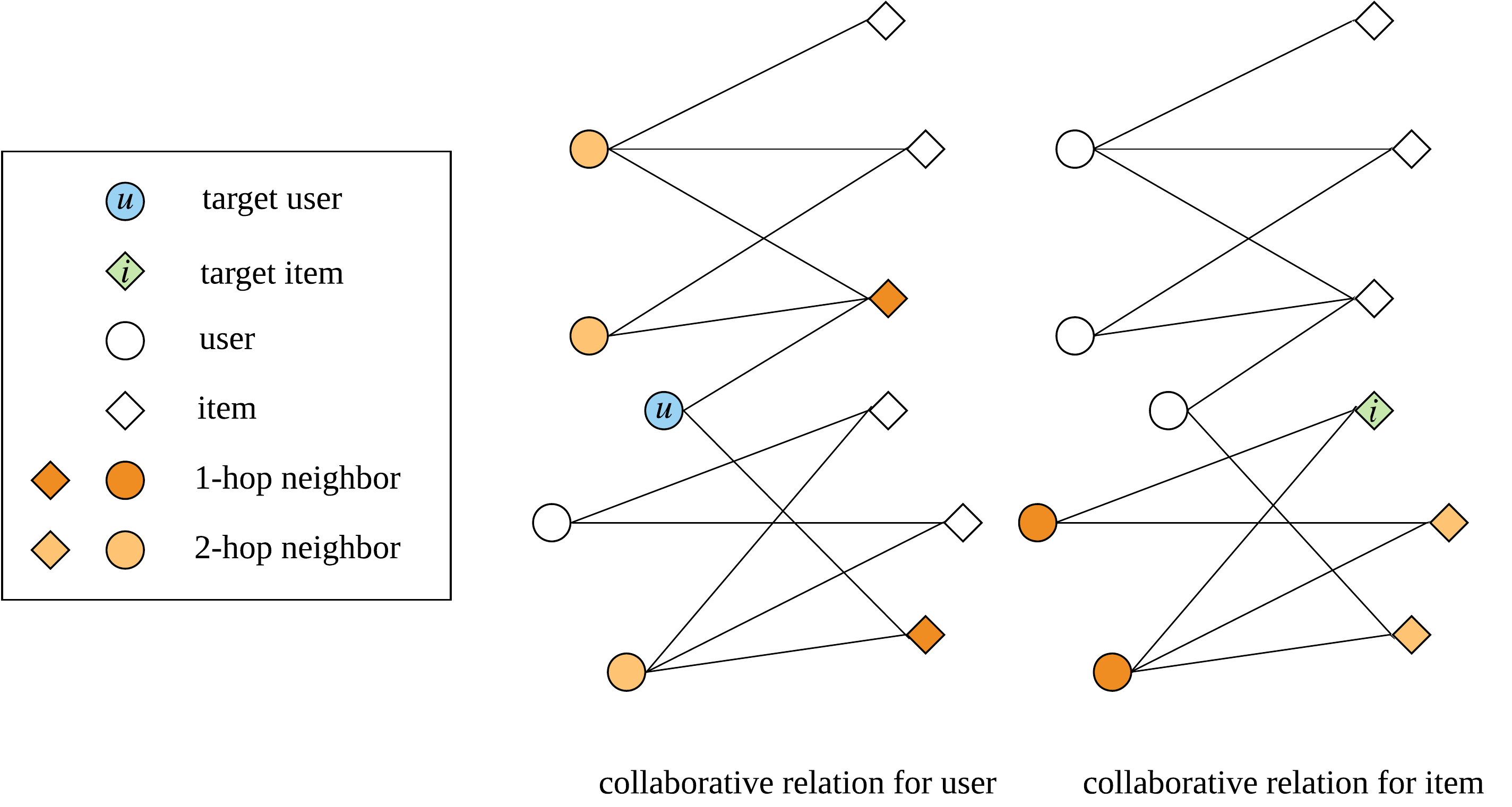}
	\caption{Graph illustration of user-item interactions.}
	\label{fig:bi-graph}
	\vspace{-10pt}
\end{figure}

We could regard the user-item interactions as a bipartite graph in which the nodes are users and items, and links are interaction records as illustrated in Figure~\ref{fig:bi-graph}. Traditional models \cite{koren2009collaborative, koren2008factorization} only consider the directly interacted items (users) of the target user (item) which are the 1-hop neighbors of the target node. In this way, it is difficult to capture the collaborative relations among users and items. But when we make a step forward and consider the 2-hop neighbors, we find that the neighbors in 2-hop have collaborative relations with the target node because both the 2-hop neighbors and the target node have interacted with the same group of nodes which are the 1-hop neighbors. As these relations are found through 2-hops on the graph, we call them \textit{high-order collaborative relations}.

By this means, we may find the corresponding collaborative information for the target user or item. Moreover, there are various patterns across the neighbors that can be utilized. By aggregating these collaborative relations to the representation of users and items, we could model more complex and various user interests (item attractions). And the collaborative modeling can be done in a sequential way to better handle the temporal dynamics.

Another key limitation of current models is that they only consider the user-side temporal dynamics while ignoring the ones on the item side. The user-side sequence consists of the items that are browsed by the user, and thus it could reveal the user's drifting interests. However, the item-side also contains sequential patterns: an item attracts different users at different time which could reveal the item dynamics such as popularity trend or social topic drift. For example, the recommender system may present Christmas card to a user when the holiday is coming even \textit{before} her interacting with any related items because the Christmas card has attracted the other users who share similar interests or collaborative relations to that specific user. The modeling of item-side sequence is similar with \textit{information dissemination} \cite{petrovic2011rt,rizoiu2018sir}, which means the item information disseminates from users to users at different time and to predict which user will be the next one that the information disseminates to.

\begin{figure}[h]
	\centering
	\includegraphics[width=1.0\columnwidth]{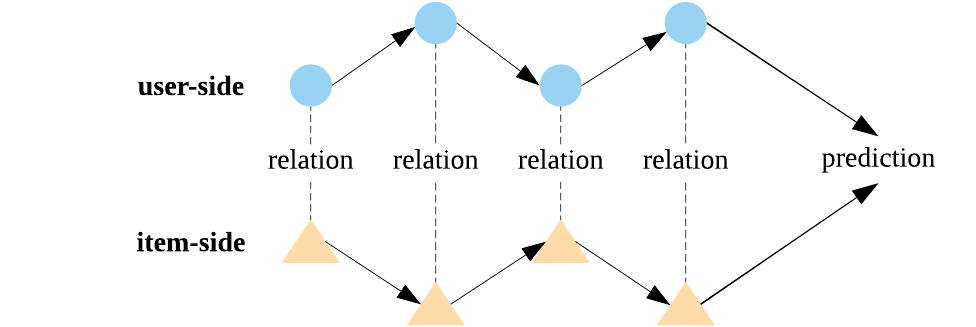}
	\caption{Illustration of interactive dual sequence modeling.}
	\label{fig:inter-dual}
\end{figure}

There are already some sequential recommendation models that have tried combining user-side and item-side sequences to perform dual sequence modeling \cite{wu2017recurrent,wu2019dual}. However, these works intend to consider the two sequences in a relatively independent manner and the sequential representations of both sequences have interactions only in the final prediction stage. 
Nevertheless, our work aims to model the dual sequences in a more interactive way which means the information of both sequences have interactions along the timeline. As we do collaborative relation capturing from both user-side and item-side, it is natural that we interact both sequences at synchronized time which is illustrated in Figure~\ref{fig:inter-dual}.

To address the limitations mentioned above, we propose \textbf{S}equential \textbf{Co}llaborative \textbf{Re}commender (\score) which considers high-order collaborative relations and models dual sequences in an interactive and more expressive manner.
The contribution of the paper can be summarized in three-fold:
\begin{itemize}[leftmargin=5mm]
	\item We propose to aggregate high-order collaborative relations which could enrich the representation of users and items. More importantly, through cross neighbor relation modeling, our model can effectively capture the various and complex patterns in the neighbor-to-neighbor collaborative relations.
	\item We propose to model both user-side and item-side sequence. Dual sequences interactions are modeled in a more thorough way, which makes the modeling of the dual sequence more expressive.
	\item We conduct extensive experiments of evaluating and comparing our model with several strong baselines over three real-world yet large-scale recommendation datasets. The results have proved the efficacy of \score ~model and the detailed analysis reveals some key principles of training our model.
\end{itemize}

The rest of the paper is organized as follows.
Section~\ref{sec:preliminaries} and Section~\ref{sec:method} present the preliminaries and describe the \score~model in detail. We also make some discussions about the model efficiency.
We conduct comprehensive experiments and present the experimental setups with the corresponding results in Section~\ref{sec:exps}.
In Section~\ref{sec:rel}, we discuss about some related works.
Finally we conclude the paper and point out some future works in Section~\ref{sec:con}.

\section{Preliminaries}\label{sec:preliminaries}
In a recommender system, there are $M$ users in $\mathcal{U}=\{u_1, \ldots, u_M\}$ and $N$ items in $\mathcal{V}=\{v_1, \ldots, v_N\}$.
In the history, any user may reveal interests on some items and the interaction behaviors would be tracked in the system as $\mathcal{Y}=\{y_{uv} | u \in \mathcal{U}, v \in \mathcal{V} \}$ and
\begin{equation}
	y_{uv} = \left\{
		\begin{array}{rcl}
			1, & & u~ \text{has interacted with} ~v; \\
			0, & & \text{otherwise.} \\
		\end{array}
	\right.
\end{equation}
The user preference $y$ is either implicit feedback \cite{agarwal2009spatio}, e.g., clicks, or explicit user rating \cite{koren2009collaborative}.
Without loss of generality, we focus on the implicit feedback which is more common in practice.
For sequential recommendation, each user-item interaction has the corresponding timestamp $ts$, thus we use the triplet $(u, v, ts)$ to denote one interaction. All the observed interaction records are denote as $\mathcal{T} = \{u, v, ts\}$.

\subsection{Interaction Relations}
For a target pair of $(u,v)$ that we need to predict the interaction probability, we could extract some interaction relations of the target user $u$ and target item $v$.

\textit{Definition 1. (Interaction Set)}:
Given the target user $u$, we can conduct the \textit{interaction set} of $u$ as
\begin{equation}
\mg(u) = \{v_j | y_{uv_j}=1 \}.
\end{equation}
The user's interaction set is the collection of all the items that the user has interacted with.
Symmetrically, we can define the item $v$'s interaction set as,
\begin{equation}
\mg(v) = \{u_i | y_{u_iv}=1 \}.
\end{equation}
which contains all the users that have interaction with the item $v$.

\textit{Definition 2. (Co-Interaction Set)}:
To explicitly capture the collaborative relations among users and items, i.e., similar users or similar items, it is natural to consider the users (items) that have similar tastes (attractions). 
Therefore we define the \textit{co-interaction set} of $u$ and $v$ respectively as
\begin{equation}
	\mg(v|u) = \{ u_i | y_{{u_i}{v_j}}=1, v_j \in \mg(u) \},\\
\end{equation}
\begin{equation}
	\mg(u|v) = \{ v_j | y_{{u_i}{v_j}}=1, u_i \in \mg(v) \}.
\end{equation}
For user $u$, the co-interaction set $\mg(v|u)$ actually consists of a group of users that shares similar behaviors with $u$, because they all interacted with the items in $u$'s interaction set. Therefore they have collaborative relations in some extent. For item $v$, similarly, the co-interaction set $\mg(u|v)$ consists of the items that attract the same group of users ($v$'s interaction set) with the target item $v$.

As shown in Figure~\ref{fig:bi-graph}, we can tell that the interaction set and co-interaction set are essentially the 1-hop and 2-hop neighbors of the rooted node ($u$ or $v$) respectively.

\subsection{Evolving Time-sliced Interaction Relations}
Now that we have defined the local interaction relations of the user $u$ (item $v$), we take one step forward and consider it in a temporal way.

Specifically, the users (items) may conduct different interactions with different items (users) at different time.
Thus, the interaction relations are evolving all the time and could be regarded as a series of time-sliced processes.

To better model the temporal patterns of the interaction relations, we slice the whole timeline into $T$ time frames, each of which is constructed within a unified time interval $\Delta T$. In this way, all observed interactions $\mathcal{T} = \bigcup_{t=1}^T \mathcal{T}^t$ and $\mathcal{T}^t$ contains the triplet $(u, v, ts)$ that happens in the $t$-th time slice.
Using the interaction records within $\mathcal{T}^t$, we could construct the user $u$'s and item $v$'s interaction relations. We denote them as,
\begin{equation}
\begin{aligned}
	\mg^t_u = \{\mg^t(u), \mg^t(v|u)\},
\end{aligned}
\end{equation}
\begin{equation}
\begin{aligned}
	\mg^t_v = \{\mg^t(v), \mg^t(u|v)\}.
\end{aligned}
\end{equation}

\subsection{Task Definition}
The goal of the recommender system is to estimate the probability of interactions $\hat{y}$ between the target user $u \in \mathcal{U}$ and the given item $v \in \mathcal{V}$, with consideration of the user's interaction history $\mg_u$ and the item's interaction history $\mg_v$ as
\begin{equation}\label{eq:pred-func}
\hat{y}_{uv} = \mathit{f}(u, v| \mg_u, \mg_v; \btheta)
\end{equation}
through the learned function $\mathit{f}$ with parameters $\btheta$ where $\mg_u = \bigcup_{t=1}^T \mg^t_u$ and $\mg_v=\bigcup_{t=1}^T \mg^t_v$.
We conclude the notations and the corresponding descriptions in Table~\ref{tab:notation}.

\begin{table}[t]
	\centering
	\caption{Notations and descriptions}\label{tab:notation}
	\resizebox{\columnwidth}{!}{
		\begin{tabular}{c|l}
			\hline
			Notation & Description. \\
			\hline
			$u, v$ & The target user and the target item. \\
			$M, N$ & The number of users and items. \\
			$y, \hat{y}$ & The indicator and the predicted probability of the user-item interaction. \\
			$\bm{u}, \bm{v}$ & Dense representation of target user $u$ and target item $v$.\\
			$\mg^t(u), \mg^t(v)$ & User $u$'s/item $v$'s interaction set at $t$-th time slice. \\
			$\mg^t(v|u), \mg^t(u|v)$ & User $u$'s/item $v$'s co-interaction set at $t$-th time slice.\\
			$\mg^t_u, \mg^t_v$ & User $u$'s/item $v$'s interaction relations at $t$-th time slice.\\
			$\bm{u}_t^{agg}, \bm{v}_t^{agg}$ & Aggregated user/item-side representation at $t$-th time slice.\\
			$U, V$ & User/item-side sequence.\\
			$\Delta T$ & Time interval to split the whole timeline.\\
			$S$ & Size of (co-)interaction set.\\
			\hline
		\end{tabular}
	}
	\vspace{-10pt}
\end{table}

\begin{figure*}[t]
	\centering
	\includegraphics[width=1.0\textwidth]{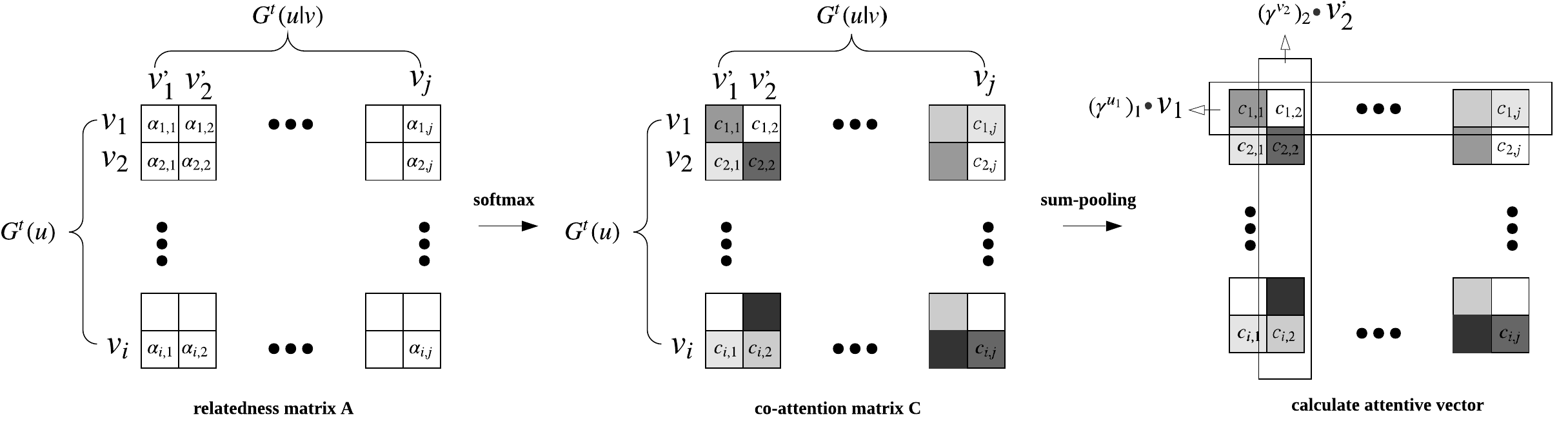}
	\caption{The processes of calculating the co-attention values. First, calculate relatedness matrix $A$; Second, softmax over the whole matrix $A$ and get the co-attention matrix $C$; Lastly, use sum pooling along rows and columns to get attentive vector and use the vector to weighted sum over objects in (co-)interaction set.}
	\label{fig:co-atten}
	\vspace{-10pt}
\end{figure*}

\section{Methodology}\label{sec:method}
In this section, we present our proposed model \score~(\textbf{S}equential \textbf{Co}llaborative \textbf{Re}commender) in detail. We first introduce the high-order collaborative relation mining through cross neighbor modeling, and then we describe the dual sequence modeling in an interactive manner. Furthermore, we analyze the time complexity of the proposed model.

\subsection{High-Order Collaborative Relation Mining}\label{sec:spatial}
In this section, we describe the proposed \textit{Co-Attention Network} for handling the complex relations across the neighbors of interaction set and co-interaction set.

\subsubsection{Cross Neighbor Co-Attention Network}
At each time slice $t$, we use \textit{Co-Attention Network} to capture the complex relations across neighbors in interaction and co-interaction sets. 

One of the key parts of recent success of recommendation models \cite{song2019session,wu2019dual1,zhou2018deepb} are the attention mechanism which attributes different credits to different item representations or temporal representations (e.g. hidden states of RNNs).
The attentive weight of a user interacted item $v_j$ w.r.t target item $v$ is calculated following the paradigm as,
\begin{equation}\label{eq:para}
	\alpha_{v_j, v} = Attn(r(v_j), v), v_j \in \mg(u),
\end{equation}
where the $Attn(\cdot)$ function can be various, $r(\cdot)$ is representation of $v_j$ which could be embedding or hidden states. The calculated $\alpha_{v_j, v}$ measures the correlation (e.g. similarity) between $v_j$ and $v$.
This paradigm only focuses on the relations between user interacted items and \textbf{single} target item $v$. But there are many neighboring items of $v$ (those items in $\mg(u|v)$), so we could calculate neighbor-to-neighbor correlations between items in $\mg(u|v)$ and those in $\mg(u)$. In this way, the relation between target $u$ and $v$ can be modeled with more resourceful information.

To model this cross neighbor collaborative relations, we propose Co-Attention Network, which is illustrated in Figure~\ref{fig:co-atten}. 
We not only consider the relatedness between the user interacted items and the target item, but also take the relatedness across user interacted items and the collaborative neighbors of the target item into account.

At each time slice, we calculate a co-attention relatedness matrix, $A^{\text{item}}_t \in R^{S \times S}$, each element of which is calculated as
\begin{equation}
	\alpha_{i, j}^t = \sigma(\bm{w}_1^T [\bm{v}_i, \bm{v}_j, \bm{v}] + b), v_i \in \mg^t(u), v_j \in \mg^t(u|v)
\end{equation}
where $\bm{v}$ is the embedding of the target item $v$, $\sigma$ is the ReLU activation function $ReLU(x) = max(0, x)$ and $S$ is the number of items in $\mg^t(u)$ and $\mg^t(u|v)$. As the objects in $\mg^t(u)$ and $\mg^t(u|v)$ are all items, we denote this relatedness matrix as $A^{\text{item}}_t$. 
Followed by a softmax operation, we get the co-attention matrix $C^{\text{item}}_t$, each element of which is calculated as,
\begin{equation}\label{eq:softmax}
	c_{i, j}^t = \frac{\exp(\alpha_{i, j}^t)}{\sum_{m,n} \exp(\alpha_{m,n}^t)}.
\end{equation}

Symmetrically, we could calculate each element of $A^{\text{user}}_t$ as
\begin{equation}
	\alpha_{i, j}^t = \sigma(\bm{w}^T_2 [\bm{u}_i, \bm{u}_j, \bm{u}] + b), u_i \in \mg^t(v), u_j \in \mg^t(v|u)
\end{equation}
and $C^{\text{user}}_t$ using the softmax operation described in Eq.~(\ref{eq:softmax}).

In this way, we could capture cross neighbor relations which are more complex and resourceful than the original paradigm described in Eq.~(\ref{eq:para}).
\subsubsection{high-order Information Aggregation}
In collaborative filtering (CF) models \cite{koren2008factorization,he2017neural,cheng2018delf}  and sequential recommendation models \cite{hidasi2015session,tang2018personalized,zhou2018deepb}, we only use the user (item) directly interacted items (users) to represent a user (item) which may cause a narrow understanding of user or item properties. 

However, in previous section, by incorporating the information of co-interaction sets, not only could we model the cross neighbor collaborative relations but integrate high-order information by summarizing the co-interaction set. By integrating these co-interaction objects, we could enrich the representation of target user $u$ and item $v$.

By sum pooling (SP) along the rows or columns of the two co-attention matrix $C^{\text{item}}_t$ and $C^{\text{user}}_t$, we could get four attentive vector,
\begin{equation}
	\bm{\gamma}_t^{u_1} = SP(\{C^{\text{item}}_{ij}\}_{j=1}^S),
\end{equation}
\begin{equation}
	\bm{\gamma}_t^{v_2} = SP(\{C^{\text{item}}_{ij}\}_{i=1}^S),
\end{equation}
\begin{equation}
	\bm{\gamma}_t^{u_2} = SP(\{C^{\text{user}}_{ij}\}_{j=1}^S),
\end{equation}
\begin{equation}
	\bm{\gamma}_t^{v_1} = SP(\{C^{\text{user}}_{ij}\}_{i=1}^S).
\end{equation}

We denote the embeddings of interaction set and co-interaction set for $u$ and $v$ at $t$-th time slice as $\bm{U}_t^1 \in R^{S \times d}$ (interaction), $\bm{U}_t^2 \in R^{S \times d}$ (co-interaction), $\bm{V}_t^1 \in R^{S \times d}$, $\bm{V}_t^2 \in R^{S \times d}$ respectively, where $d$ is the dimension of the embedding and $S$ is the size of (co-)interaction set.
The aggregated representation of $u$ and $v$ at $t$-th time slice are
\begin{equation}
	\bm{u}^{agg}_t = [{\bm{U}_t^1}^T\bm{\gamma}_t^{u_1}, {\bm{U}_t^2}^T\bm{\gamma}_t^{u_2}]
\end{equation}
\begin{equation}
	\bm{v}^{agg}_t = [{\bm{V}_t^1}^T\bm{\gamma}_t^{v_1}, {\bm{V}_t^2}^T\bm{\gamma}_t^{v_2}].
\end{equation}

\subsection{Interactive Dual Sequence Modeling} \label{sec:temporal}

In this section, we describe our approach on temporal dynamics modeling. We conduct a dual sequence modeling method which considers both user-side and item-side sequences and interactively models the relations among two sequences of synchronized time slice.

\subsubsection{Temporal Dynamics Modeling}
At each time slice, we get the aggregated representations of target user $u$ and target item $v$ as $\bm u_t^{agg}$ and $\bm v_t^{agg}$ respectively following the co-attention mechanism in Section~\ref{sec:spatial}.
After that we get two sequences $U$ and $V$, which are the sequences of target user's and item's aggregated representation  at different time respectively. For simplicity, we denote $\bm u_t = \bm u_t^{agg}$ and $\bm v_t = \bm v_t^{agg}$ thus
\begin{equation}
U = \{\bm{u}_1, \bm{u}_2, ..., \bm{u}_{T}\}~,
V = \{\bm{v}_1, \bm{v}_2, ..., \bm{v}_{T}\} .
\end{equation}

We use two recurrent neural network models to model the temporal dynamics for user-side and item-side respectively. And we implement each recurrent cell as Gated Recurrent Unit \cite{cho2014learning} (GRU).
Each GRU unit takes the corresponding representation $\bm u_t$ (or $\bm v_t$) at each time step and the hidden state $\bh_{t-1}$ from the last time step, and then calculates as
\begin{equation}\small
\begin{aligned}
\bm{z}_t^u =& ~\sigma (\overline{\bm{W}}_z^u \bm{u}_t + \overline{\bm{U}}_z \bh_{t-1}^u + \overline{\bm{b}}_z^u) \\
\bm{r}_t^u =& ~\sigma (\overline{\bm{W}}_r^u \bm{u}_t + \overline{\bm{U}}_r \bh_{t-1}^u + \overline{\bm{b}}_r^u) \\
\bh_{t}^u =& ~(1 - \bm{z}_t^u) \odot \bh_{t-1}^u  \\
&~~~~~~~~~~~~~~~~~+ \bm{z}_t^u \odot \tanh(\overline{\bm{W}}_h^u \bm{u}_t + \overline{\bm{U}}_h^u (\bm{r}_t^u \odot \bh_{t-1}^u) + \overline{\bm{b}}_h^u)  ~,
\end{aligned}
\label{eq:gru}
\end{equation}
where $\odot$ is the element-wise product operator. 

The item-side temporal dynamics are modeled in the same way. Till now, we've got user-side and item-side sequence of temporal representations: $\{\bm h_1^u, \bm h_2^u, ..., \bm h_T^u\}$ and $\{\bm h_1^v, \bm h_2^v, ..., \bm h_T^v\}$.

\subsubsection{Interactive Attention Mechanism of Dual Sequence} \label{sec:attention}
\begin{figure}[t]
	\centering
	\includegraphics[width=1.0\columnwidth]{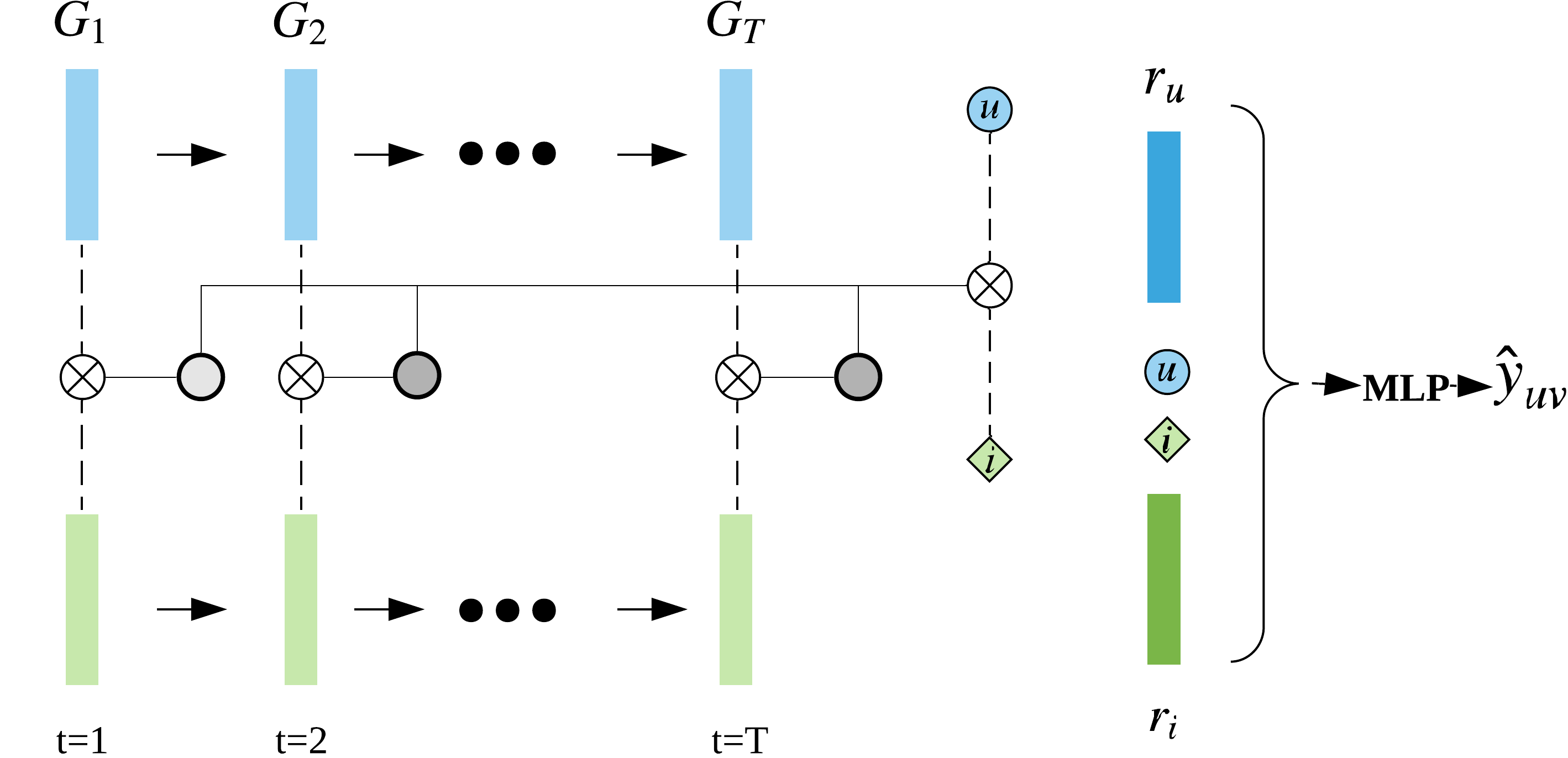}
	\caption{Temporal Interactive Dual Sequence Modeling of SCoRe.}
	\label{fig:temporal-framework}
	\vspace{-10pt}
\end{figure}

As illustrated in Figure~\ref{fig:temporal-framework}, different time slice has different impact on the final prediction at $(T + 1)$. And hereby we introduce our \textit{Interactive Attention Mechanism}. Unlike the attention mechanism in \cite{zhou2018deepb} and \cite{zhou2018deepa} which uses the target item to query the interacted items sequence, we utilize dual sequences information at the same time interactively to weigh across different time slice.
The attention value of each time slice $\beta_t$ is calculated as,
\begin{equation}
relation_t = R([\bm h_t^u, \bm h_t^v, \bm{\gamma}_t^{u_1}, \bm{\gamma}_t^{v_2}, \bm{\gamma}_t^{u_2}, \bm{\gamma}_t^{v_1}], [\bm u, \bm v])
\end{equation}

\begin{equation}
\beta_t = \frac{\exp(relation_t)}{\sum_{i=1}^T \exp(relation_i)}
\end{equation}
where R is a three-layer MLP with ReLU activation function.

And the final representations of user-side and item-side sequences are 
\begin{equation} \label{eq:seq_rep}
\bm r_u = \sum_{t=1}^T \beta_t \bm h_t^u, ~~~~\bm r_v = \sum_{t=1}^T \beta_t \bm h_t^v
\end{equation}

It is natural that we consider both side information to calculate attention, because the cross neighbor relations modeling described in Section~\ref{sec:spatial} utilizes both user-side neighbors and item-side neighbors. As a result, we consider using both sides representations of synchronized time slice to interactively calculate attentive value. In this way, the modeling of two sequences are highly correlated.

\subsection{Final Prediction and Loss Functions}
The predicted probability of interaction between the target user and the target item is calculated as
\begin{equation}
\hat{y} = f(\bm{r}_u, \bm{r}_v, \bm{u}, \bm{v}; \bm{\Theta}),
\end{equation}
where $f$ is implemented as a multi-layer perceptron with the ReLU activation function. The parameters set of the MLP is $\bm{\Theta}$.
The inference procedure is illustrated in Figure~\ref{fig:temporal-framework}.

As for the loss function, we take an end-to-end training and introduce (i) the widely used cross entropy loss $\mathcal{L}_{\text{ce}}$ \cite{ren2018bid,zhou2018deepa,zhou2018deepb} over the whole training dataset and (ii) the parameter regularization $\mathcal{L}_{\text{r}}$. We utilize Adam algorithm for optimization. Thus the final loss function is
\begin{equation}
\begin{aligned}
\argmin_{\bm{\Theta}, \bm{\Phi}} &= \mathcal{L}_{\text{ce}} + \lambda \mathcal{L}_r \\
&= -\sum_{s} \big[ y_s \log \hat{y}_s + (1-y_s) \log (1-\hat{y}_s)\big] \\
& + \frac{1}{2} \lambda \left( \|\bm{\Theta} \|_2^2 + \| \bm{\Phi} \|_2^2 \right) ~,
\end{aligned}
\end{equation}
where $\bm{\Phi}$ includes the parameters in GRUs, $\bm{w}_1$, $\bm{w}_2$ in Co-Attention Network and the parameters of the three-layer MLP $R$ in the Interactive Attention Mechanism.

\subsection{Model Efficiency}
In this section, we analyze the computational complexity of our \score~model.
From the previous sections, we can tell that the forward inference of \score~can be regarded as two relatively separate parts. 
The first part is the cross neighbor collaborative relations modeling, which can be paralleled conducted for each time slice. The cost of it can be viewed as a constant $O(C_{co-atten})$ as the co-attention network conduct single layer non-linear transformation, softmax and sum-pooling operations. 
The second part is the GRU temporal modeling. We assume the average time performance of the GRU is a constant $O(C_{GRU})$ which is related to the implementation of the GRU module yet can be parallelly executed through GPU processor. Recall that we have $T$ time slices, thus the time complexity of temporal inference is $O(T \cdot C_{GRU})$. 
Therefore the overall time complexity of \score~is $O(C_{co-atten}) + O(T \cdot C_{GRU}) = O(T \cdot C_{GRU})$ which is the time complexity of ordinary recurrent neural networks.

The inference time complexity is acceptable since several implementations sharing similar execution complexity have been adopted online \cite{wu2016personal,zhou2018deepb}, which indicates online inference efficiency for our \score~model in some extent.

\section{Experiments}\label{sec:exps}
In this section, we present the details of the experiment setups and the corresponding results. To illustrate the effectiveness of our proposed model, we compare it with some strong baselines on sequential recommendation task. Moreover, we have published our reproductive code\footnote{https://github.com/qinjr/SCoRe}.

We start with three research questions (RQ) to lead the experiments and the following discussions.
\begin{itemize}
	\item [\textbf{RQ1}] Compared to the baseline models, does SCoRe achieve state-of-the-art performance in sequential recommendation task?
	\item [\textbf{RQ2}] What is the influence of different components in~\score? Are the proposed co-attention network and interactive attention necessary for improving performance?
	\item [\textbf{RQ3}] What patterns does the proposed model capture for the final recommendation decision?
\end{itemize}

\subsection{Experimental Setups}
In this part, we describe our experiment setups including datasets with preprocessing method, some important implementation details, evaluation metrics and the compared baselines.
\subsubsection{Datasets}
We use three real-world large-scale datasets to evaluate all the compared models. The dataset statistics have been shown in Table ~\ref{tab:dataset-statistics}.
\begin{description}[leftmargin=15pt]
	\item [CCMR] \cite{cao2016complete} is a dataset consists of movie rating (from integer score 1 to 5) logs collected from Douban, which is one of China's largest movie review websites. The data is collected and dumped in May 2015.
	\item [Taobao] \cite{zhu2018learning} is a dataset consisting of user behavior data collected from Taobao\footnote{https://tianchi.aliyun.com/dataset/dataDetail?dataId=649}, one e-commerce platform in China. It contains user behaviors from November 25 to December 3, 2017 of several behavior types including click, purchase, add to cart and item favoring.
	\item [Tmall] \footnote{https://tianchi.aliyun.com/dataset/dataDetail?dataId=42} is provided by Alibaba Group which contains user behavior history on Tmall e-commerce platform from May 2015 to November 2015.
\end{description}

\noindent\textbf{Dataset Preprocessing}.
We cut the time line into total $T$ time slices with the specific time interval as shown in Table~\ref{tab:dataset-statistics}. And for each time slice, we use the interaction records within it to construct interaction and co-interaction set for both user and item. Here we use a simple way to do time slicing, we leave finer segmentation strategy in future work.

\begin{table}[t]
	\centering
	\caption{The dataset statistics.}\label{tab:dataset-statistics}
	\resizebox{\columnwidth}{!}{
		\begin{tabular}{c|r|r|r|r|r}
			\hline
			Dataset & Users \# & Items \# & Interaction \# & Time slices $T$ & Time interval $\Delta T$\\
			\hline
			CCMR & 4,920,695 & 190,129 & 283,775,314 & 41 & 90 days\\
			\hline
			Taobao & 987,994 & 4,162,024 & 100,150,807
			 & 9 & 1 day\\
			\hline
			Tmall & 424,170 & 1,090,390 & 54,925,331 & 13 & 15 days\\
			\hline
		\end{tabular}
	}
\end{table}

\noindent\textbf{Positive \& Negative Samples}.
To evaluate the recommendation performance, we use one positive item and sample 100 negative items at the prediction time $(T + 1)$ for each user in all three datasets.
For Tmall and Taobao datasets, as we only have the positive user feedbacks (click, buying, etc.), we have to randomly sample the negative items. As for CCMR datasets, we regard items whose ratings are 5 or 4 as positive items and those whose ratings are 1,2 or 3 as negative items. If a user does not have enough negative items, we use random sampling to generate negative items for her. The positive items in CCMR form the behavior sequence.

\noindent\textbf{Train \& Test Splitting}.
The training set contains the sequential behaviors from the first to the $(T-2)$th time slice, we use the interactions history from 1 to $(T-3)$ to predict in $(T-2)$. For the validation set, we use the interactions data from 1 to $(T-2)$ to predict in $(T-1)$. In testing set, interactions data from 1 to $(T-1)$ are used to predict in $T$.

\noindent\textbf{Implementation Details}.
It is common that the target user doesn't have any interaction record in a time slice, and similarly, the target item may be not visited by any user in a time slice. To handle this issue, we use a unified embedding vector to represent the situation.

We set the size of interaction set to $S$ which can be regarded as a hyperparameter. For simplicity, the size of co-interaction set is $S$ too.
If there are more than $S$ objects in a set, we use random sampling. If there are less than $S$ objects (say $k$ ($<S$) objects), we random sample $(S-k)$ times among the original set.

\subsubsection{Evaluation Metrics}
Three evaluation metrics are used and all of them are widely used in recommendation tasks.

\textbf{HR@k} (\textit{Hit Ratio@k}) measures the proportion of samples that the positive item is among the top-k in all test cases which is computed as,
\begin{equation}
HR@k = \frac{1}{|\mathcal{U}|} \sum_{u \in \mathcal{U}} \mathbb I (R_{u, v} \leq k),
\end{equation}
where $R_{u,v}$ is the ranking position of the user $u$'s interacting with item $v$, and $\mathbb I$ is the indicator function.

\textbf{NDCG@k} (Normalized Discounted Cumulative Gain) is a position-aware metric which assigns larger weights on higher ranks of the positive item, which is calculated as,
\begin{equation}
NDCG@k = \frac{1}{|\mathcal{U}|} \sum_{u \in \mathcal{U}} \frac{2^{\mathbb I (R_{u, v} \leq k)} - 1} {log(R_{u, v} + 1)}.
\end{equation}

\textbf{MRR} (Mean Reciprocal Rank) is another position-aware metric that is calculated as,
\begin{equation}
MRR = \frac{1}{|\mathcal{U}|} \sum_{u \in \mathcal{U}} \frac{1}{R_{u, v}}.
\end{equation}

As HR@1 is equal to NDCG@1, so in this work, we report HR@\{1,5,10\}, NDCG@\{5,10\} and MRR in detail.

\subsubsection{Compared Baselines}\label{sec:comp-models}
To illustrate the effectiveness of our model, we compare \score ~with two CF models, three single sequence recommendation models and two dual sequence models.
We follow \cite{zhou2018deepb} that all the models take the input sparse features and feed them through an embedding layer for the subsequent inference.

The first group of models are CF models:
\begin{itemize}[leftmargin=40pt]
	\item [\textbf{SVD++}] \cite{koren2008factorization} is a hybrid method of latent factor model and neighbor-based model which is the fundamental approach of collaborative filtering recommendation. It regards all the sequential behaviors as a whole and ignores the temporal dynamics.
	\item [\textbf{DELF}] \cite{cheng2018delf} is the state-of-the-art CF method which utilizes deep neural networks to capture complex non-linear interaction patterns from both user-side and item-side.
\end{itemize}

The second group contains sequential recommendation methods that utilize single user-side sequence, which are based on RNNs, CNNs, or Transformer architecture:
\begin{itemize}[leftmargin=40pt]
	\item [\textbf{GRU4Rec}] \cite{hidasi2015session} bases on GRU and it is the first work using the recurrent cell to model sequential user behaviors.
	\item [\textbf{Caser}] \cite{tang2018personalized} is based on CNNs that uses horizontal and vertical convolutional filters to capture user behavior patterns at different scales.
	\item [\textbf{SASRec}] \cite{kang2018self} bases on Transformer \cite{vaswani2017attention}, it only uses self-attention mechanism without recurrent architecture. It achieves very competitive performance in sequential recommendation task.
\end{itemize}

The third group is dual sequence recommendation models.
\begin{itemize}[leftmargin=40pt]
	\item [\textbf{RRN}] \cite{wu2017recurrent} is the first RNN-based model that considers both the user- and item-side sequence. It uses sum-pooling to aggregate the information inside a time slice.
	\item [\textbf{DEEMS}] \cite{wu2019dual} feed the user-side and item-side sequence respectively into two identical sequential models, and let the two models play a game with each other where one model will use the predicted score of the other model as feedback to guide the training.
	\item [\textbf{SCoRe}] is our proposed model which is described in Section~\ref{sec:method}.
\end{itemize}

\begin{table*}[h]
	\centering
	\caption{Performance comparison against baseline models. Bold values are the best in each row, while the second best values are underlined. Improvements over baselines are statistically significant with $p$ < 0.01. (HR, NDCG, MRR: the higher, the better)}\label{tab:perf-table}
	\tiny
	\resizebox{0.9\textwidth}{!}{
		\begin{tabular}{c|c|cc|ccc|ccc}
			\hline
			\multirow{2}{*}{Dataset} & \multirow{2}{*}{Metric} & \multicolumn{2}{c|}{Group 1} & \multicolumn{3}{c|}{Group 2} & \multicolumn{3}{c}{Group 3}\\
			& & SVD++ & DELF & GRU4Rec & Caser & SASRec & RRN & DEEMS & \score\\
			\hline
			\multirow{6}{*}{CCMR} & HR@1 & 0.0797 & 0.0755 & 0.0739 & 0.0845 & 0.0817 & 0.0739 & \underline{0.0968} & \textbf{0.1035} \\
			& HR@5 & 0.1865 & 0.2255 & 0.2477 & 0.2469 & \underline{0.2480} & 0.2214 & 0.2444 & \textbf{0.2518} \\
			& HR@10 & 0.2686 & 0.3422 & 0.3494 & \underline{0.3663} & 0.3613 & 0.3431 & 0.3599 & \textbf{0.3688} \\
			& NDCG@5 & 0.1340 & 0.1638 & 0.1689 & 0.1736 & \underline{0.1779} & 0.1733 & 0.1776 & \textbf{0.1891} \\
			& NDCG@10 & 0.1604 & 0.2051 & 0.1985 & 0.2113 & \underline{0.2128} & 0.2060 & 0.2115 & \textbf{0.2167} \\
			& MRR & 0.1516 & 0.1750 & 0.1706 & 0.1829 & 0.1893 & 0.1799 & \underline{0.1896} & \textbf{0.1954} \\
			\hline
			\hline
			\multirow{6}{*}{Taobao} & HR@1 & 0.1947 & 0.3381 & 0.3439 & \underline{0.3562} & 0.3510 & 0.3204 & 0.3255 & \textbf{0.3688} \\
			& HR@5 & 0.4489 & 0.6077 & 0.6035 & 0.6085 & 0.6159 & 0.6220 & \underline{0.6478} & \textbf{0.6816} \\
			& HR@10 & 0.5933 & 0.7084 & 0.7189 & 0.7224 & 0.7371 & \underline{0.7620} & 0.7517 & \textbf{0.8068} \\
			& NDCG@5 & 0.3256 & 0.4731 & 0.4866 & 0.5005 & \underline{0.5101} & 0.4779 & 0.4814 & \textbf{0.5339} \\
			& NDCG@10 & 0.3723 & 0.5089 & 0.5139 & 0.5174 & 0.5199 & 0.5233 & \underline{0.5476} & \textbf{0.5745} \\
			& MRR & 0.3224 & 0.4405 & 0.4617 & 0.4744 & 0.4818 & 0.4615 & \underline{0.4988} & \textbf{0.5121} \\
			\hline
			\hline
			\multirow{6}{*}{Tmall} & HR@1 & 0.3447 & 0.3386 & 0.3501 & 0.3588 & 0.3622 & 0.3634 & \underline{0.3669} & \textbf{0.3770} \\
			& HR@5 & 0.5594 & 0.5636 & 0.5727 & 0.5712 & 0.5819 & 0.7310 & \underline{0.7331} & \textbf{0.7381} \\
			& HR@10 & 0.6554 & 0.6562 & 0.6646 & 0.6662 & 0.6686 & \underline{0.8378} & 0.8373 & \textbf{0.8479} \\
			& NDCG@5 & 0.4589 & 0.4654 & 0.4784 & 0.4768 & 0.4843 & \underline{0.5594} & 0.5565 & \textbf{0.5693} \\
			& NDCG@10 & 0.4901 & 0.4986 & 0.5080 & 0.5074 & 0.5124 & 0.5942 & \underline{0.5951} & \textbf{0.6051} \\
			& MRR & 0.4527 & 0.4669 & 0.4741 & 0.4730 & 0.4778 & 0.5256 & \underline{0.5259} & \textbf{0.5363} \\
			\hline
		\end{tabular}
	}
	\vspace{-10pt}
\end{table*}

\subsection{Evaluation Results: RQ1} \label{sec:comp_results}

The experimental results are shown in Table~\ref{tab:perf-table}, we find several observations as below.

By comparing the performance of \score~ and other baseline models, it outperforms baselines by 28.9\% to 3.1\%, 58.8\% to 2.7\% and 18.5\% to 2.0\% on MRR in CCMR, Taobao and Tmall dataset, respectively. And it also shows significant improvements on the other metrics so \score~ achieves the state-of-the-art performance in sequential recommendation task.

For the models in Group 1, they do not consider the temporal dynamics of user behaviors thus perform not so good as models in Group 2 and 3. DELF uses both user-side and item-side interaction information so it achieves better performance than SVD++ which only utilizes user-side information.

By comparing the performance of Group 2 and 3, we find in Tmall and Taobao dataset that, Group 3 outperforms Group 2. But over CCMR dataset, SASRec and Caser are better than Group 3. As shown in Table~\ref{tab:dataset-statistics}, Tmall and Taobao has a lot more items than CCMR, which makes ranking items more difficult in Tmall and Taobao. So it is more important on these two datasets to take item-side sequence into consideration because it gives the models more information than just using single user-side sequence.

DEEMS performs better than RRN in many cases which means the two player game of dual sequence in DEEMS is effective and show the potential of finer design on dual modeling.

\begin{figure}[t]
	\centering
	\includegraphics[width=1.0\columnwidth]{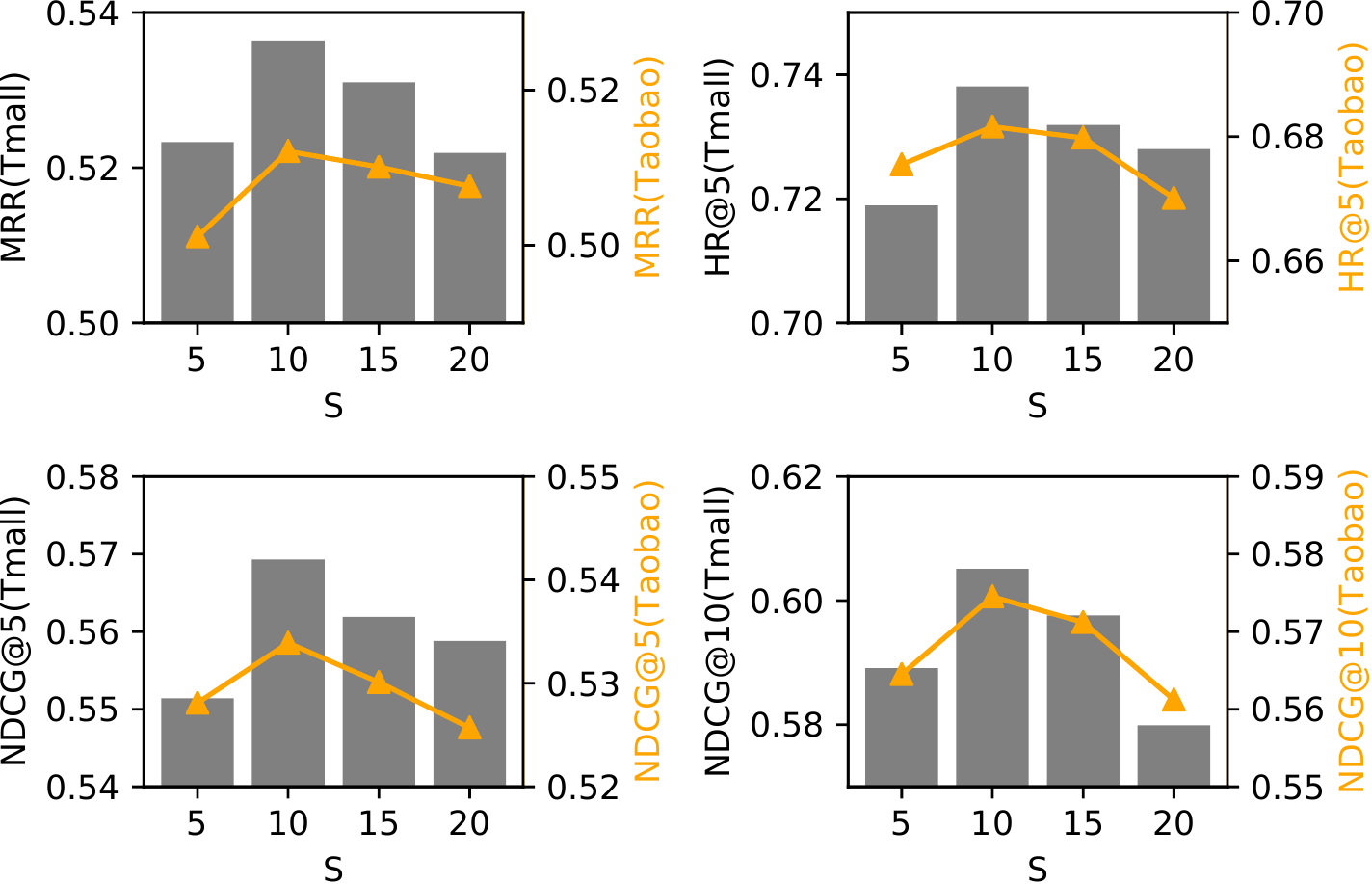}
	\caption{Performance comparision on different size $S$ of (co-)interaction set on Tmall and Taobao dataset.}
	\label{fig:s-study}
	\vspace{-10pt}
\end{figure}

\textbf{Influence from the size of (co-)interaction set}.
We vary the size of (co-)interaction set to further investigate the robustness of \score. For simplicity, we set interaction and co-interaction set to have same size in dual side. The results on Tmall and Taobao dataset are shown in Figure~\ref{fig:s-study}. 
We find that when size increases, the performance is improved at first because that the larger the size is, the more information it contains. And when the size continues to increase, the performances begin to drop which indicates that too much noise and useless information is introduced.

\subsection{Ablation Study: RQ2} \label{sec:ab-study}
In this section, we conduct some ablation studies to investigate the effectiveness of three important components of \score~: (1) Interactive Attention Mechanism in dual sequence modeling; (2) Co-Attention Network for cross neighbor relation mining and aggregation; (3) The consideration of using both user-side and item-side.

\begin{table}[h]
	\centering
	\caption{Performance comparison of ablation study.} \label{tab:ab_study_table}
	\resizebox{\columnwidth}{!}{
		\begin{tabular}{c|c|ccccc}
			\hline
			\multirow{2}{*}{ Dataset } & \multirow{2}{*}{ Metric } & \multicolumn{5}{c}{ models }\\
			& & RIA & RCA & User & Item & \score~ \\
			\hline
			\multirow{3}{*}{ CCMR } & HR@10 & 0.3667 & 0.3461 & 0.3491 & 0.3218 & \textbf{0.3688} \\
			& NDCG@10 & 0.2098 & 0.2077 & 0.2012 & 0.1892 & \textbf{0.2167} \\
			& MRR & 0.1872 & 0.1795 & 0.1782 & 0.1567 & \textbf{0.1954} \\
			\hline
			\multirow{3}{*}{ Taobao } & HR@10 & 0.7888 & 0.7702 & 0.7678 & 0.7453 & \textbf{0.8068} \\
			& NDCG@10 & 0.5436 & 0.5286 & 0.5192 & 0.5001 & \textbf{0.5745} \\
			& MRR & 0.4785 & 0.4669 & 0.4652 & 0.4495 & \textbf{0.5121} \\
			\hline
			\multirow{3}{*}{ Tmall } & HR@10 & 0.8467 & 0.8406 & 0.8355 & 0.8122 & \textbf{0.8479} \\
			& NDCG@10 & 0.6030 & 0.5903 & 0.5843 & 0.5671 & \textbf{0.6051} \\
			& MRR & 0.5352 & 0.5289 & 0.5192 & 0.5085 & \textbf{0.5363} \\
			\hline
		\end{tabular}
	}
	\vspace{-10pt}
\end{table}

We set four comparative settings, and the performances of them have been shown in Table~\ref{tab:ab_study_table}.
The details of the four settings are listed as below.
\begin{itemize}[leftmargin=15pt]
	\item \textbf{RIA} (Remove Interactive Attention) removes the attention part described in~\ref{sec:attention} and set the final sequential representations of target user and item as $\bm r_u = h_T^u$ and $\bm r_v = h_T^v$ of Eq.~(\ref{eq:seq_rep}).
	\item \textbf{RCA} (Remove Co-Attention) removes the co-attention network and uses simply sum pooling to aggregate neighbor information in interaction and co-interaction set.
	\item \textbf{User} only uses the user-side sequence to do final prediction, as $\hat{y} = f(\bm{r}_u, \bm{u}, \bm{v}; \bm{\Theta})$.
	\item \textbf{Item} only uses the item-side sequence to do final prediction, as $\hat{y} = f(\bm{r}_v, \bm{u}, \bm{v}; \bm{\Theta})$.
\end{itemize}
Except for the changes mentioned above, the other parts of the models and experimental settings remain identical to ensure the fairness of comparison.

From Table~\ref{tab:ab_study_table} we can find that (1) \score~performs the best indicating the efficacy of different components of the model. (2) the performance decreases more when removing co-attention than interactive attention which means the cross neighbor relation modeling is more important and fundamental to \score's performance. (3) Using single sequence hurt the performance badly and item-side is harder to model compared to user-side thus have worse performance.

\subsection{Case Study: RQ3}
\begin{figure}[h]
	\centering
	\includegraphics[width=1.0\columnwidth]{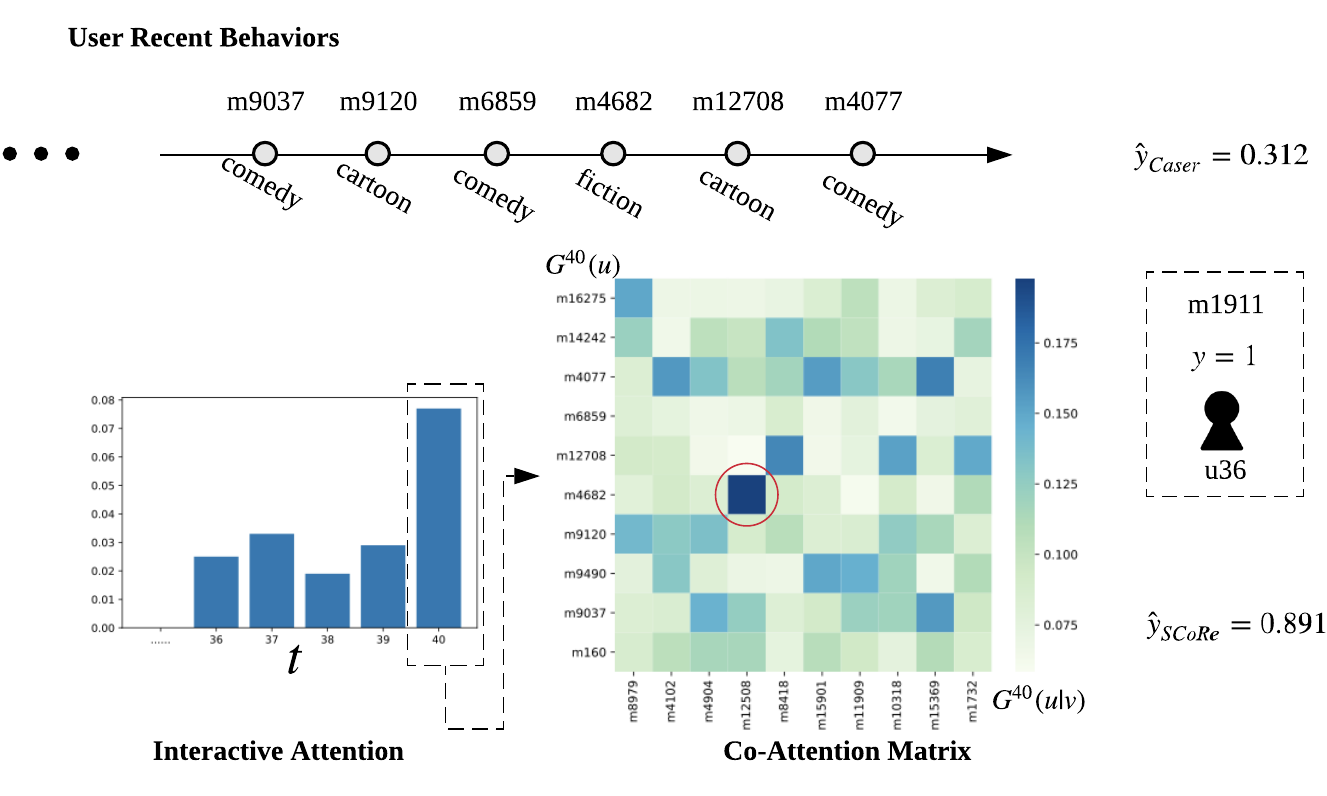}
	\caption{Case study of Caser and \score. We plot user u36's recent behaviors, interactive attention values and co-attention matrix $C^{\text{item}}_{40}$ of time slice 40 of \score.}
	\label{fig:case_study}
	\vspace{-10pt}
\end{figure}

In this section, we further investigate what patterns \score~captures by studying and visualizing a specific case sampled from the CCMR dataset. 
In Figure~\ref{fig:case_study}, we plot the prediction of user-item pair (u36, m1911) where m1911 is the movie \textit{American Sniper}. The ground truth is $y=1$, we plot the Caser and \score~ predictions which are $\hat{y}_{\text{caser}}=0.312$ and $\hat{y}_{\text{\score}}=0.891$ respectively. By looking into the user behavior sequence, we find the reason of \score's better prediction result.

The user's recent behaviors are favouring comedy, cartoon or fiction, as illustrated in the upper part of Figure~\ref{fig:case_study} which are not very relevant to the target item \textit{American Sniper} which is a biographical and  action movie. So it is natural that models like Caser tend to give lower prediction score.

However, we plot the interactive attention value of recent time slices of \score~and find the 40-th slice have higher attention value. So we further plot the co-attention matrix of this time slice $C^{\text{item}}_{40}$. We find m4682 (\textit{Interstellar}, fiction and adventure) of $\mg^{40}(u)$ has high relation with m12508 (\textit{007:Casino Royal}, action and fiction) of $\mg^{40}(u|v)$. Movie m12508 is more relevant to the user's interest, and its representation is aggregated to the item-side, so it is reasonable that \score~has the ability to give the target item higher predicted probability score, which has precisely been interacted by the target user in the test data.

\section{Related Work}\label{sec:rel}
\subsection{Collaborative Filtering}
In recommender system literatures, the most widely used method is collaborative filtering \cite{goldberg1992using}, which learns from the historical user-item interactions without exogenous information about items or users. It recommends according to the modeled user preferences, e,g., clicks \cite{qu2016product,agarwal2009spatio} and ratings \cite{koren2009collaborative}, over the items.
Many works \cite{koren2009matrix,salakhutdinov2007probabilistic,yang2012local,zhang2013optimizing} have been proposed based on collaborative filtering.
Among them, latent factor models play a key role in CF methods, ranging from pLSA \cite{hofmann2004latent} and Latent Dirichlet Allocation \cite{blei2003latent} to SVD-based models \cite{koren2008factorization,chen2012svdfeature} and factorization machines \cite{rendle2010factorization}.
Nowadays, deep neural network (DNN) has attracted huge attentions in recommender systems because of its effective feature extraction and end-to-end model training with satisfying generalization \cite{zhang2017deep}.
Some DNN methods \cite{he2017neural,he2018outer,qu2016product} are proposed for latent factor collaborative filtering.
However, almost all of these approaches, either conventional matrix factorization methods or deep models, lack of temporal pattern mining. 

\subsection{Sequential Recommendation}

Recently, sequential recommendation has drawn huge attention since the sequences of user behaviors have rich information for the user interests, especially for concept drifting \cite{widmer1996learning}, long-term behavior dependency \cite{koren2009collaborative,ren2019lifelong}, periodic patterns \cite{ren2018repeatnet}, etc.

There are three categories for sequential recommendation.
The first one is from the view of temporal collaborative filtering \cite{koren2009collaborative} with the consideration of drifting user preferences.
The second stream is based on Markov-chain methodology \cite{rendle2010factorizing,he2016fusing,he2016vista} which implicitly models the user state dynamics and derive the outcome behaviors.
The third school is based on deep neural networks, such as recurrent neural networks (RNNs) \cite{hidasi2015session,hidasi2017recurrent,wu2017recurrent,jing2017neural,liu2016context,beutel2018latent,villatel2018recurrent} and convolutional neural networks (CNNs) regarding the behavior history as an image \cite{tang2018personalized,kang2018self}.
However, most of these methods only care about user's interest drifting and do not consider the sequential patterns of items, which also deliver rich information for user-item matching. Models like \cite{wu2017recurrent, wu2019dual} considers both sequences but in a relatively independent way, which leaves space for finer design of dual sequence modeling.
Furthermore, most of these sequential models only care about user's own interaction history while ignoring the information that could be found in similar users or items. And thus the sequential models may suffer from narrowness of recommendation.

\section{Conclusion and Future Work}\label{sec:con}
In this paper, we propose \score, a model that utilizes and aggregates high-order collaborative information using cross neighbor modeling to improve representation learning and collaborative relation mining.
Furthermore, we propose an interactive attention mechanism to model the user-side and item-side sequences. 
In this way, dual sequence modeling captures temporal dynamics from both user and item-side and significantly facilitate final recommendation performance.

For the future work, we plan to further investigate on the time segmentation strategy of the evolving sequential interactions and its influence to the recommendation performance. We also seek to deploy our method on the real-world recommender systems.

\textbf{Acknowledgement}. The corresponding author Weinan Zhang thanks the support of National Natural Science Foundation of China (61702327, 61772333, 61632017) and Shanghai Sailing Program (17YF1428200).

\bibliographystyle{ACM-Reference-Format}
\balance
\bibliography{wsdm725}

\end{document}